\documentclass[11pt]{article}

\usepackage[preprint]{acl}

\usepackage{times}
\usepackage{latexsym}

\usepackage[english]{babel}
\usepackage{amsmath}
\usepackage{amsfonts}
\usepackage{amssymb}
\usepackage{comment}
\usepackage{float}
\usepackage{caption}
\usepackage{color}
\usepackage{algorithmic}
\usepackage{algorithm}
\usepackage{multirow}
\usepackage{bold-extra}
\usepackage{fancyhdr}
\usepackage{siunitx}
\usepackage{titling}
\usepackage{placeins}
\usepackage{float}

\usepackage{fancyvrb,fvextra}
\usepackage{dblfloatfix}
\usepackage{enumitem}
\usepackage{subcaption}
\definecolor{c0}{HTML}{1f77b4}
\definecolor{c1}{HTML}{ff7f0e}
\definecolor{c2}{HTML}{2ca02c} %
\definecolor{c3}{HTML}{d62728} %
\definecolor{c4}{HTML}{9467bd} 
\definecolor{c5}{HTML}{8c564b}
\definecolor{c6}{HTML}{e377c2}
\definecolor{c7}{HTML}{7f7f7f}
\definecolor{c8}{HTML}{bcbd22}
\definecolor{c9}{HTML}{17becf}

\usepackage{booktabs}
\usepackage{makecell} %
\usepackage{rotating}

\usepackage{pdflscape}
\usepackage{capt-of}   %

\usepackage{tabularx}
\usepackage{tablefootnote}

\usepackage{array}

\usepackage{dblfloatfix}

\newcolumntype{Y}{>{\RaggedRight\arraybackslash}X}

\usepackage{ragged2e}

\usepackage{hyperref}
\usepackage{shellesc}

\usepackage{amsmath}

\usepackage[T1]{fontenc}
\usepackage[utf8]{inputenc}

\usepackage{microtype}

\usepackage{hyperref}

\usepackage{inconsolata}

\usepackage{graphicx}
\usepackage{appendix}

\title{Why Large Language Models can Secretly Outperform Embedding Similarity in Information Retrieval}

\author{Matei Benescu \\
  University of Groningen \\
  \texttt{matei.benescu22@gmail.com} \\\And
  Ivo Pascal de Jong \\
  University of Groningen \\
  \texttt{ivo.de.jong@rug.nl} \\}

\begin{document}
\maketitle

\begin{abstract}

With the emergence of Large Language Models (LLMs), new methods in Information Retrieval are available in which relevance is estimated directly through language understanding and reasoning, instead of embedding similarity. We argue that similarity is a \textit{short-sighted} interpretation of relevance, and that LLM-Based Relevance Judgment Systems (LLM-RJS) (with reasoning) have potential to outperform Neural Embedding Retrieval Systems (NERS) by overcoming this limitation. Using the TREC-DL 2019 passage retrieval dataset, we compare various LLM-RJS with NERS, but observe no noticeable improvement. 
Subsequently, we analyze the impact of reasoning by comparing LLM-RJS with and without reasoning. We find that human annotations also suffer from short-sightedness, and that false-positives in the reasoning LLM-RJS are primarily mistakes in annotations due to short-sightedness. We conclude that LLM-RJS do have the ability to address the short-sightedness limitation in NERS, but that this cannot be evaluated with standard annotated relevance datasets.

\end{abstract}

\section{Introduction}\label{sec:introduction}

Information Retrieval (IR) is the task of finding relevant information in a pool of numerous documents, where a document can be any retrievable unit like a webpage or text fragment. %
IR methods have constantly been evolving (\citealp[]{kazi2025bridging}). Vector Space Models (\citealp[]{salton1965evaluation, salton1975vector}) use sparse vector representations of term weights, where each weight represents the frequency of the term in the given document. Documents are ordered by relevance to the query, by the means of an ordering function, such as the BM25 (\citealp[Chapter 3.4]{robertson2009probabilistic}). Modern IR methods rely on dense vector representations of text, known as ``embeddings", which are learned by Neural Networks (\citealp[]{huang2013learning}). We will refer to these methods as Neural Embeddings Retrieval Systems (NERS). In NERS, documents are ordered by their similarity with the query, which is given by a distance metric such as Euclidean Distance, or Cosine Similarity (\citealp[]{wang2020measurement}). \citealp{karpukhin2020dense} showed that NERS outperform sparse vector representation approaches.

\begin{figure}[t]
    \centering
    \includegraphics[width=\linewidth]{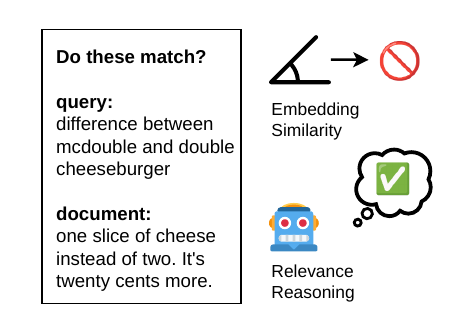}
    \caption{Motivation for LLM-RJS. The query and document are not similar, so embedding similarity is too short-sighted for IR. The document is the correct answer, but requires reasoning about relevance to identify as a match.}
    \label{fig:concept_relevance_reasoning}
\end{figure}

More recently, Large Language Models (LLMs) have been proposed as a different approach to judging how relevant a document is to a query (\citealp[]{macavaney2023one, faggioli2023perspectives, abbasiantaeb2024can}). \citealp{alaofi2024llms} showed that LLMs' relevance judgments are similar to those of humans. This has the potential to outperform NERS models, as NERS models are limited by retrieving documents that are \textit{similar} to the query, but relevant documents are not always similar. An example of NERS being short-sighted is shown in Figure \ref{fig:concept_relevance_reasoning}.

For the scope of this paper, we will call a system that uses an LLMs verbalized relevance score for matching an \textit{LLM-Based Relevance Judgment System} (LLM-RJS). While LLM-RJS have been shown to be effective \cite{alaofi2024llms}, to the best of our knowledge a 1-to-1 comparison against NERS is missing.
Additionally, IR applications of reasoning models that are equipped with Chain of Thought (CoT) capabilities, such as GPT-o1 (\citealp[]{jaech2024openai}) remain unexplored. As reasoning capabilities become cheaper and available in open models (\citealp[]{guo2025deepseek}), examining its impact on document relevance judgments becomes increasingly important and broadly applicable.
We aim to build on \citealp{alaofi2024llms} by comparing LLM-RJS to NERS. Furthermore, we explore the effects of reasoning capabilities of LLMs in the task of document relevance estimation.

With this, we make the following contributions:
\begin{itemize}[noitemsep, topsep=0pt]
    \item We evaluate NERS and multiple LLM-RJS models on the TREC-DL 2019 passage retrieval benchmark to compare their performances in relevance ranking. We see that top LLMs perform on par with NERS, but do not yield demonstrably better rankings.
    \item We evaluate the impact of ``reasoning effort'' for LLM-RJS at directly estimating the relevance scores. We propose that this can help to overcome the short-sightedness of NERS. We find that adding reasoning results in higher relevance ratings that the human relevance scores, and that the LLM-RJS overall shows higher relevance ratings. 
    \item We explain the difference in relevance ratings as short-sightedness in the ground-truth annotations, showing examples of documents that have low relevance ratings in the human annotation, but that are highly relevant. From the 94 `false positives', 89 are actually mistakes in the annotation because the passages are relevant but not similar.
    \item We conclude that the empirical capabilities of zero-shot methods for Information Retrieval may be underestimated due to short-sighted annotations, making short-sighted similarity models appear better than more capable reasoning models. LLM-RJS have the ability to overcome the short-sighted limitation of NERS, but this does not lead to better metrics when evaluated on human annotations. 

\end{itemize}

\begin{table*}[ht]
\centering
\footnotesize
\setlength{\tabcolsep}{4pt}
\begin{tabularx}{\textwidth}{
@{}
>{\RaggedRight\arraybackslash}p{0.22\textwidth}
X
c
@{}
}
\toprule
\textbf{Query} & \textbf{Passage} & \textbf{Rel.} \\
\midrule

axon terminals or synaptic knob definition &
At the end of the axon's branches and form synapses with the dendrites of other neurons. & 0 \\
\addlinespace[0.6em]

definition declaratory judgment &
That is especially true in situations where a non-practicing entity has sent a licensing offer to a potential infringer. Cases finding declaratory judgment jurisdiction. A patent owner wrote to its licensee alleging that a product sold by the licensee was subject to a license agreement. The licensee paid the requested royalties under protest and filed a declaratory judgment suit. & 1 \\
\addlinespace[0.6em]

difference between a mcdouble and a double cheeseburger &
Although we previously reported that the McDouble would be on the Dollar Menu, in Manhattan the cheeseburger appears on the Dollar Menu while the McDouble is \$1.39 and the double cheeseburger is \$1.49. & 2 \\
\addlinespace[0.6em]

do goldfish grow &
A: The conditions goldfish are kept in plus their diet determine how large they will grow. I have seen goldfish grow ridiculously large in very small containers when their water was changed frequently. Goldfish will not grow if water conditions are poor. Fancy goldfish don't grow as large as Common goldfish. A good size would be around 5 inches body length for most fancy varieties, 8 inches for Comets and 12 inches for Common Goldfish. These sizes are usually only attained by pond grown fish. & 3 \\

\bottomrule
\end{tabularx}
\caption{Examples from TREC-DL 2019 test set. Relevance: 0 = Irrelevant, 1 = Related, 2 = Highly relevant, 3 = Perfectly relevant.}
\label{tab:triples_examples}
\end{table*}

\section{Related Work}

We describe previous research on NERS and LLM-RJS models. While NERS stem from a long line of existing research in text embeddings \cite{mikolov2013efficient, pennington2014glove}, we specifically focus on embeddings from Neural Networks.

\paragraph{NERS}

Neural Embedding Retrieval Systems (NERS) rely on pre-trained language models to create a dense vector representation of text. These language models are Neural Networks trained on some other such as Masked Language Modeling \citep{devlin2019bert} or Contrastive Learning \cite{gao2021simcse}. The penulatimate layer of the Neural Network therefore has a dense representation of the semantic meaning of the text, capturing properties like synonymy \cite{karpukhin2020dense}. 

NERS models work by computing the embedding representation of the query and the document, and evaluating the distance between them, typically through cosine similarity. When a document and a query describe similar concepts they get a higher similarity score. Document relevance is therefore ranked by similarity to the query. 

We consider this a fundamental limitation of IR systems that rely on embedding similarities. While similarity between query and document is a good heuristic for relevance, it is not exactly the same. When the document answers the question but does not repeat content from the query (as the example in Figure \ref{fig:concept_relevance_reasoning}), it is not similar, but can still be highly relevant. Even with ever-improving embedding models \cite{trabelsi2021neural}, this necessarily gives a limitation of NERS.

\paragraph{LLM-RJS}

Large Language Model-Based Relevance Judgement Systems (LLM-RJS) instead use the generated answers from an LLM to establish relevance. In this work, the LLM is prompted with query and document, and instructed to quantify how ``relevant'' the document is in line with the human annotations, though many variations for prompting exist \cite{arabzadeh2025benchmarking}.  

\citealp{faggioli2023perspectives} found that LLMs can perform comparably to humans in identifying irrelevant information but struggled to consistently recognize relevant pieces of information. In their results GPT-3.5 showed a higher false negative rate than humans. 

Evaluating the goodness of document relevance assigned by LLMs can be directly compared to human-labeled query-document pairs are used as the ``ground truth". However, humans are not consistent in rating document relevance (\citealp[]{10.1145/1390334.1390447}). To evaluate LLMs fairly, \citealp{alaofi2024llms} posed the issue as ``if
 LLMs produce labels that agree with humans to the same extent
 that humans agree with each other (human-to-human range), we can conclude that they are
 sufficiently reliable for use". Their results show that larger LLMs, such as GPT-4, GPT-4o, and LLaMA70B give results in the human-to-human range. Only GPT-4 and GPT-4o were robust to the prompt variations. However, this perspective cannot show when LLMs might outperform humans in the ability to estimate relevance.

 Because LLM-RJS systems are zero-shot models and are not trained as supervised learning systems on the human annotations, we should consider that it has the potential to outperform human annotations. Additionally, with reasoning models \cite{wei2022chain}, we can generate reasoning processes that show relevance in a way that has more depth than simple similarity scores, and might have more depth than the human labels.

\section{Dataset}

TREC (Text REtrieval Conference) datasets are commonly used in the IR field (\citealp{macavaney2023one, faggioli2023perspectives, alaofi2024llms}). Similarly to \citealp[]{alaofi2024llms}, we use queries and passages from the test set of the the passage retrieval task dataset of the Deep Learning Track of TREC 2019 (TREC-DL 2019) (\citealp[]{trec2019}). TREC-DL 2019 contains triples of form (query, passage, relevance score). 

The passage corpora of TREC-DL 2019 are derived from the MS MARCO dataset. MS MARCO is a large-scale dataset of real web search queries and web text originally sampled from Bing search logs. Queries in MS MARCO are real user queries and passages are extracted from web documents obtained via Bing search results. Table \ref{tab:triples_examples} gives some examples of (query, passage, relevance) triples. %

The relevance scores of TREC-DL 2019 are evaluated by human assessors. The scores (\citealp[Section 2.2]{trec2019}) are integer values using a 4-point scale (0-3) where:
\begin{itemize}[noitemsep, topsep=0pt]
    \item[3.] \textbf{Perfectly relevant:} The passage is dedicated to the query and contains the exact answer.
    \item[2.] \textbf{Highly relevant:}  The passage has some answer for the query, but the answer may be a bit unclear, or hidden amongst extraneous information.
    \item[1.]  \textbf{Related:} The passage seems related to the query but does not answer it.
    \item[0.] \textbf{Irrelevant:} The passage has nothing to do with the query.
\end{itemize}
For both experiments, the LLMs are instructed to only output the relevance score of the (query, document) pair. Any response that could not be casted to an integer between 0 and 3 was considered a failed response. All others are successes. For each model, the (query, document) pairs were divided into successes and failures. The intersection of all successful pairs was kept for final analysis.

\section{Experiment 1}
\label{Experiment1}

\begin{figure*}[t]
    \centering
    \includegraphics[width=\textwidth]{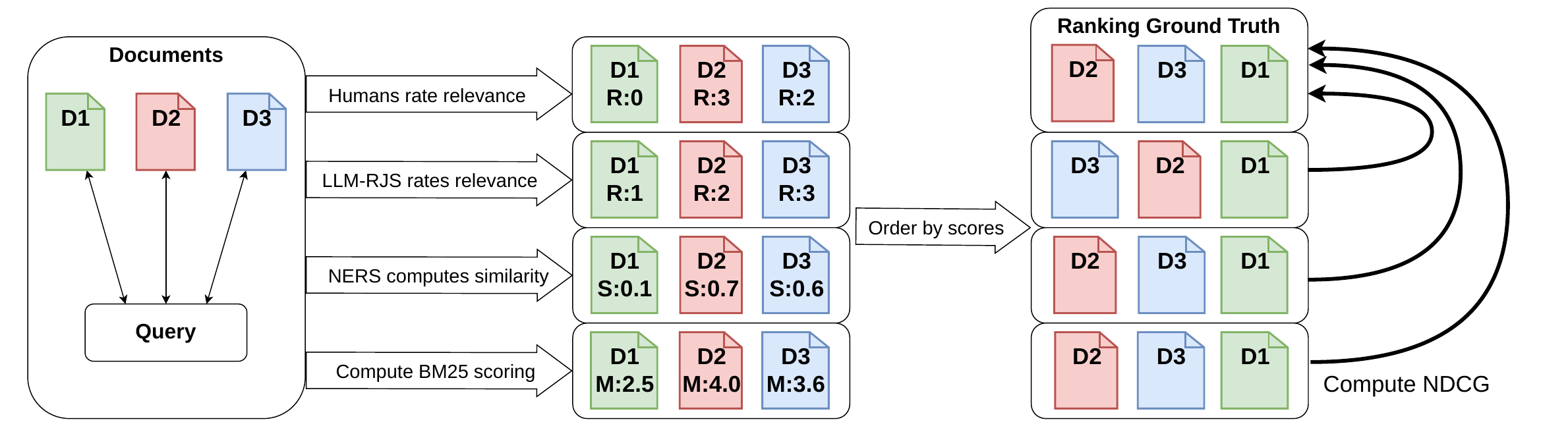}
    \caption{Overview of Experiment 1. Documents are ranked (ordered) based on the "relevance score" with respect to the query assigned by each system. The rankings are compared to the ranking yielded by humans labelers, via NDCG. The mean NDCG is used as final evaluation.}
    \label{fig:overview}
\end{figure*}

To evaluate whether LLM-RJS can outperform NERS, and therefore attempt to address the short-sightedness issue of NERS, we compare the performance of NERS with LLM-RJS in document ranking. As a baseline we evaluate the BM25 model. We first describe the implementation details of the three models, and then how they are evaluated. The overall setup of Experiment 1 is demonstrated in Figure \ref{fig:overview}.

\paragraph{LLM-RJS} All models were asked to estimate the relevance of (query,
passage) pairs on a 4-point scale pair, using the prompt found in Appendix~\ref{Prompt}. The prompt describes the task and provides the relevance score convention. The models were also given examples of (query, document, relevance) triples, as examples are known to enhance LLM performance \cite{brown2020language}. We use both closed state-of-the-art (at the moment of writing), as well as open-weights models. Table~\ref{tab:llms} provides an overview of the LLM models evaluated. The temperature hyperparameter \citep{ackley1985learning, peeperkorn2024temperature}, controls the amount of ``randomness'' in LLM token generation. The temperature was set to 0 for all models, to ensure reproducibility and because higher temperatures have been shown to have a moderate correlation with incoherence (\citealp[]{peeperkorn2024temperature}).
The reasoning efforts of GPT-5.1 was set to ``none''. The open-weights models were run using Ollama\footnote{https://ollama.com/}. 

\begin{table}[t]
\centering
\small
\setlength{\tabcolsep}{6pt}
\begin{tabular}{@{}l l c@{}}
\toprule
\textbf{Model} & \textbf{Provider} & \textbf{Open Weights} \\
\midrule
GPT-5.1         & OpenAI    & No  \\
Sonnet 4.5      & Anthropic & No  \\
Gemini 2.5 Pro  & Google    & No  \\
gpt-oss:20b     & OpenAI    & Yes \\
gpt-oss:120b    & OpenAI    & Yes \\
llama3.2:1b     & Meta      & Yes \\
llama3.2:3b     & Meta      & Yes \\
llama3.1:8b     & Meta      & Yes \\
\bottomrule
\end{tabular}

\caption{LLMs used in Experiment~1 \cite{openai_gpt51, anthropic_sonnet45, google_gemini25, ollama_gptoss, meta_llama32, meta_llama31}.}

\label{tab:llms}
\end{table}

\begin{figure*}[t]
\centering
\begin{subfigure}[t]{.45\linewidth}
  \centering
  \includegraphics[height=5.3cm]{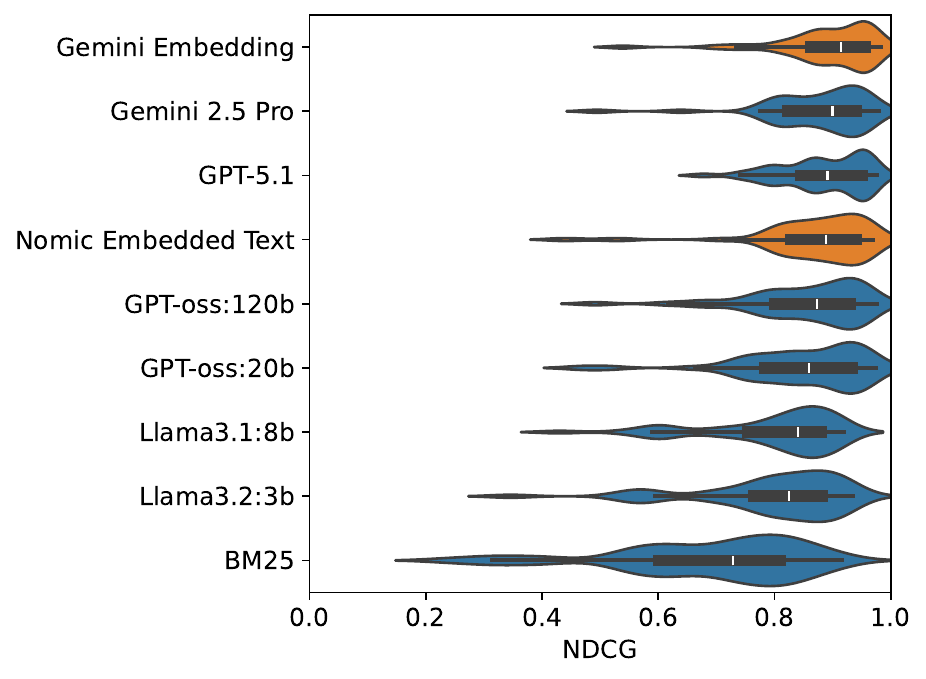}
\end{subfigure}\hfill
\begin{subfigure}[t]{.27\linewidth}
  \centering
  \includegraphics[height=5.3cm,]{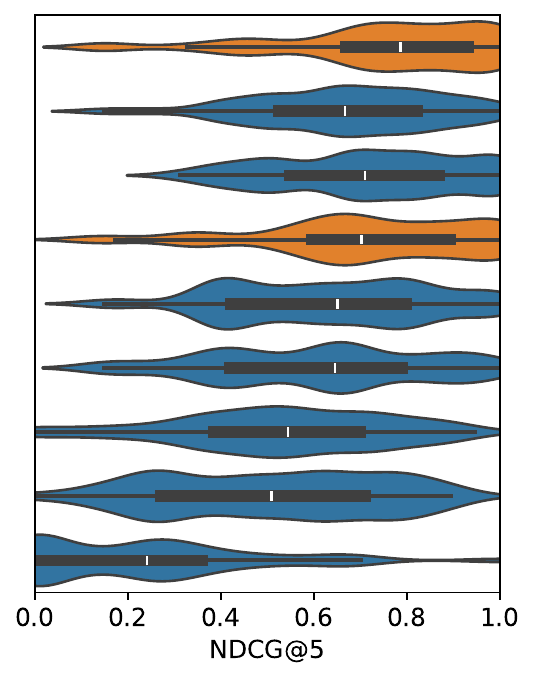}
\end{subfigure}\hfill
\begin{subfigure}[t]{.27\linewidth}
  \centering
  \includegraphics[height=5.3cm]{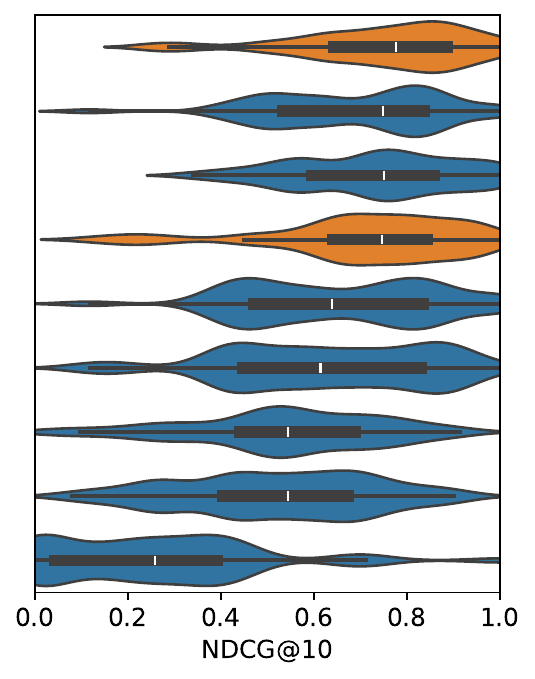}
\end{subfigure}\hfill
\caption{Performance of \textcolor{c0}{\textbf{LLM-RJS}} and \textcolor{c1}{\textbf{NERS}} models in relevance ranking, measured by NDCG@k. Top LLM-RJS models perform on par with NERS models, but do not appear to outperform them. Results are in line with previous research from \citet{hofstatter2021efficiently, xing2023query}, reporting mean NDCG@10 of 0.72 and 0.73 respectively.}
\label{fig:violin_plots}
\end{figure*}

\paragraph{Neural Embeddings Retrieval Systems:}
For NERS, we used a closed model and an open-weights one. The closed model, gemini-embedding001, is the state-of-the-art from Google, showed to outperform most of the current embedding models (\citealp[]{lee2025gemini}). The open model is nomic-embed-text:v1.5\footnote{https://huggingface.co/nomic-ai/nomic-embed-text-v1.5}. It is an extension of nomic-embed-text:v1 (\citealp[]{nussbaum2024nomic}) that uses Matryoshka Representation Learning (\citealp[]{NEURIPS2022_c32319f4}). The base model, nomic-embed-text:v1, showed better performance  (\citealp[]{nussbaum2024nomic}) than similarly sized embedding models and established models like GloVe (\citealp[]{pennington2014glove}). 

\paragraph{BM25 baseline:} BM25 is a probabilistic scoring function that ranks documents based on query term frequencies while balancing the effects of term frequency and document length. 

The BM25 scoring function is defined as:
\[
\resizebox{\columnwidth}{!}{$
\mathrm{BM25}(d,q)=
\sum_{t\in q}\mathrm{IDF}(t)
\frac{f_{t,d}(k_1+1)}
{f_{t,d}+k_1\left(1-b+b\frac{|d|}{\mathrm{avgdl}}\right)}
$}
\]
where \(q\) is the query, \(f_{t,d}\) is the frequency of term \(t\) in document \(d\), \(k_1\) controls the degree of term-frequency saturation, \(|d|\) is the length of document \(d\), \(\text{avgdl}\) is the average document length in the collection, and \(b \in [0,1]\) controls the strength of document length normalization.

The inverse document frequency (IDF) reduces the weight of common terms while increasing the importance of rarer terms across the collection:
\[
\text{IDF}(t) = \log \frac{N - n_t + 0.5}{n_t + 0.5}
\]
where \(N\) is the total number of documents and \(n_t\) is the number of documents containing term \(t\).

Similarly to \citealp[]{trec2019}, BM25 with the default parameters ($k_1=1.5$, $b=0.75$) was chosen as a baseline ranking system.
The rankings yielded by both LLM-RJS and NERS will be compared to the one of BM25.

\paragraph{Model evaluation:}

The relevance estimates from NERS and BM25 cannot be directly compared to ground truth relevance scores, due to different representations of relevance. Instead, the document rankings, i.e. the set of documents ordered by their relevance to the query, yielded by the different systems can be directly compared. \\The Normalized Discounted Cumulative Gain (NDCG) (\citealp[]{wang2013theoretical}) is a measure of ranking quality, which we will use to compare the rankings obtained by different retrieval systems. \\NDCG is defined as:
\[
\mathrm{NDCG}(A)=\frac{\mathrm{DCG}(A)}
{\mathrm{DCG}(I)},
\]\[\quad
\mathrm{DCG}(R)=\sum_{i=1}^{N}\frac{h(d_i)}{\log(i+1)},
\]
where \(I\) is the ideal ranking and \(h(d_i)\) is the relevance of document $d$ at position $i$ in the ranking. NDCG ranges from 0 to 1, with 1 corresponding to the ideal ranking induced by human relevance judgments.\\
NDCG@k is NDCG evaluated for the sub-ranking that consists of the first $k$ documents in the initial ranking.

Since each query yields a ranking, and the dataset consists of multiple queries, the average NDCG and NDCG@k with $k=5$ and $k=10$ over queries will be computed as final evaluation metrics. 
\subsection{Results and Discussion}
For each query, all the passages in relation with it have been ranked, and the NDCG, NDCG@5 and NDCG@10 has been computed for the rank yielded by each model. Figure~\ref{fig:violin_plots} shows the results of Experiment 1 as a violin plot with NDCG, NDCG@5 and NDCG@10 of all models. %

The NERS and LLM-RJS models performed on par with previous models \citealp[]{hofstatter2021efficiently, xing2023query}. These previous models are purposely trained dense retrievers, unlike the models analyzed by this paper, which are zero-shot. Note that the zero-shot models seem to outperform trained retrievers. The presented NDCG@10 of these previous models was reported in the respective papers and not implemented here. %

For this experiment, only Llama3.1:8b, Llama3.2:1b and Sonnet 4.5 had any failures, out of which, Llama3.2:1bs and Sonnet 4.5 have been excluded for having failed more than 2/3 of all data points (lack of instruction compliance, see Appendix~\ref{Prompt}. Table~\ref{tab:failed_examples} provides an example of failure for each model.). The models failed to only output one integer between 0 and 3. Expectedly, one of the excluded models is Llama3.2:1b, the smallest model in the selection, with 8312/9260 failed responses. Surprisingly, the other one was Sonnet 4.5, the state-of-the-art (at the time of writing) of Anthropic, which failed to only give relevance scores for 7473/9260 data points. Llama3.1:8b only failed 84/9260 data points.

\begin{figure*}[t]
    \centering
    \includegraphics[width=\textwidth]{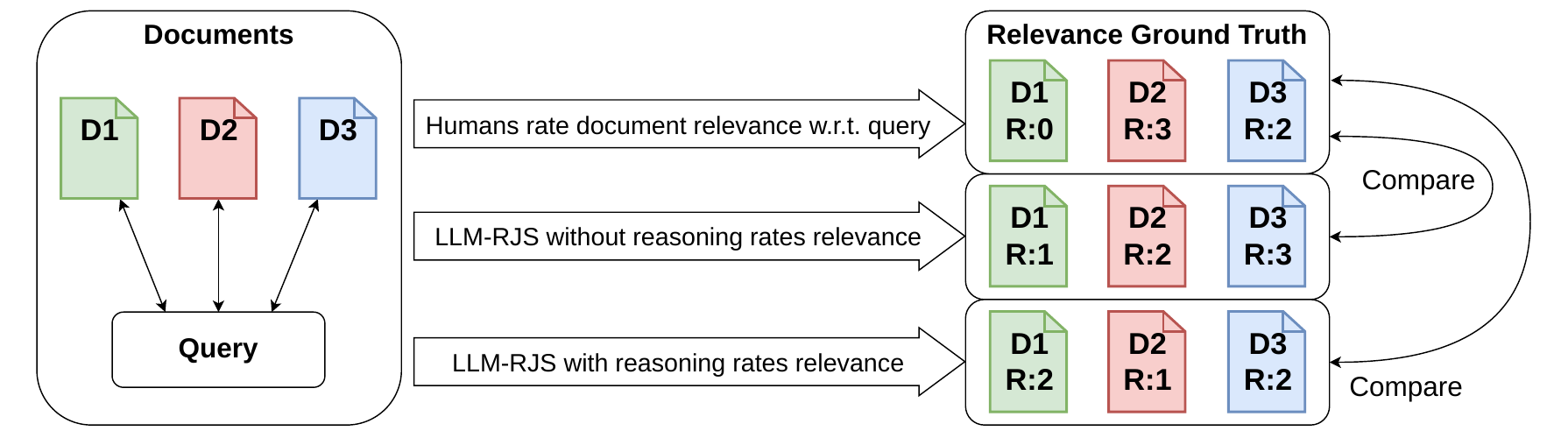}
    \caption{Overview of Experiment 2. The LLM estimates relevance of (query, document) pairs in "high reasoning" and "low reasoning" conditions. The mean absolute error between LLM and human assessment in the two conditions is used as evaluation metric. %
    }\vspace{0.5em}
    \label{fig:overview_reasoning}
\end{figure*}

Overall, NERS are still most proficient for ranking documents, and perform the task better for only a fraction of the cost, and a fraction of the time. The price per 1 million tokens strongly favors NERS, with Gemini Embeddings costing 0.15\$ per 1 million tokens. On the other side, GPT-5.1 costs 1.25\$ per 1 million input tokens and 10\$ per 1 million output tokens. Albeit financially undesirable, LLMs do not leave much on the table, when it comes to ranking (query, document) pairs.

The closed-source models consistently outperform the open models. The bottom line is, if one is looking for the best option, Gemini Embeddings is the winning model. If, aside performance, running the retrieval system locally is also a consideration, nomic-embed-text:v1.5 is a viable option. %

\section{Experiment 2}\label{Experiment2}

\begin{table*}[t]
\centering
\footnotesize
\setlength{\tabcolsep}{3pt}
\renewcommand{\arraystretch}{1.1}
\begin{tabularx}{\textwidth}{@{}
>{\RaggedRight\arraybackslash}p{0.18\textwidth}
>{\RaggedRight\arraybackslash}X
c c c c
@{}}
\toprule
\textbf{Query} & \textbf{Passage} & \textbf{Human} & \textbf{Reasoning} & \textbf{No Reasoning} & \textbf{NERS} \\
\midrule

axon terminals or synaptic knob definition & The small bulges at the end of the axon that send messages to other neurons are called & 0 & 3 & 2 & 0.48\\\addlinespace[0.6em]

causes of left ventricular hypertrophy & Heart hypertrophy is a condition where the muscular wall of the heart becomes thickened. [...] It can thicken as a result of high blood pressure [...] & 0 & 3 & 2 & 0.52 \\\addlinespace[0.6em]

cost of interior concrete flooring & Polished concrete pricing and cost per square metre for concrete polishing can range from as little as £20 to as much as £150 per metre square [...] & 0 & 3 & 2 & 0.46 \\\addlinespace[0.6em]

definition declaratory judgment & n. a judgment of a court which determines the rights of parties without ordering anything be done or awarding damages [...] & 0 & 3 & 3 & 0.53 \\\addlinespace[0.6em]

difference between a mcdouble and a double cheeseburger	& One slice of cheese instead of two. It's twenty cents more to get a double cheeseburger for \$1.19. [...] & 0 & 3 & 3 & 0.43 \\\addlinespace[0.6em]

difference between rn and bsn & Once you graduate, you have your BSN, but to become an RN, you have to take state boards and pass them before you are an RN. [...] & 0 & 3 & 3 & 0.57 \\\addlinespace[0.6em]

\textbf{do goldfish grow} & a\textbf{verage ones are four to six inches (10 -15 cm) long, but they can grow up to 7 inches} [...] & 0 & 3 & 2 & \textbf{0.35} \\\addlinespace[0.6em]

\textbf{example of monotonic function} & [...] \textbf{Since then, this proportion has grown steadily and monotonically with each successive election to 43\% in 1996, 54\% in 2000, 58\% in 2004} [...] & 0 & 3 & 2 & \textbf{0.14} \\\addlinespace[0.6em]

exons definition biology & A portion of DNA that codes for a section of the mature messenger RNA obtained from that DNA [...] & 0 & 3 & 2 & 0.55\\\addlinespace[0.6em]

\textbf{how long is life cycle of flea} & [...] \textbf{Adults live about 7 to 10 days and deposit eggs on the moist soil surface or in soil cracks} [...] & 0 & 3 & 2 & \textbf{0.38}\\\addlinespace[0.6em]

is cdg airport in main paris & It is located near Roissy, 25 km to the north-east of Paris. The construction of this airport lasted about 10 years [...] & 0 & 3 & 2 & 0.47\\

\bottomrule
\end{tabularx}
\caption{%
First 10 query-IDs where GPT5.1 gave rating 3 while the humans gave 0. The annotations suffer from short-sightedness, while GPT-5.1 with reasoning correctly identifies relevance. Of the 94 such cases only 5 were genuine false positives (shown in Appendix \ref{ap:false_positives}). The NERS  similarity averages 0.44 for these 94 cases, while it is 0.50 for true positives (higher is more similar). \textbf{Bold} highlights key examples of the short-sightedness limitation.}
\label{tab:examples_reasoning_nuance}
\end{table*}

Chain of Thought (CoT) reasoning \cite{wei2022chain} has been shown to yield substantial improvement in LLM capabilities on a variety of tasks. It works by letting the LLM generate a longer series of tokens as ``reasoning'' before giving the final answer. This strategy has been shown to lead to better results in image Information Retrieval \cite{wu2025chain}, and is therefore also promising for having LLM-RJS outperform NERS models. 

We prompt GPT-5.1 to estimate the relevance of (document, query) pairs with the prompt described in Appendix \ref{ap:appendixPrompt}. To compare the reasoning against the non-reasoning conditions we change the ``reasoning effort''\footnote{https://platform.openai.com/docs/guides/reasoning\#get-started-with-reasoning}, which controls the amount of reasoning tokens before the output. A visual overview of the experiment is shown in Figure \ref{fig:overview_reasoning}. 

The same experiment was also performed with gpt-oss:20b, a smaller open-source model, with similar results reported in Appendix \ref{ap:oss-reasoning}.

\begin{figure}[H]
\centering
\begin{subfigure}[t]{.54\linewidth}
  \centering
  \includegraphics[height=4.2cm]{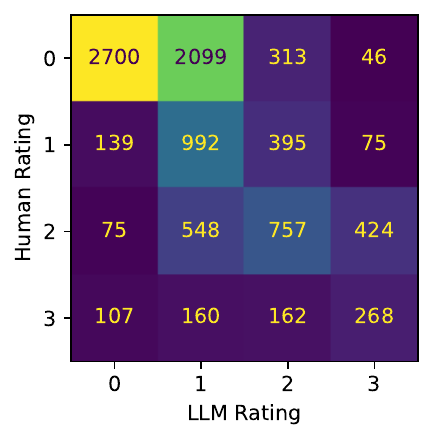}
\end{subfigure}\hfill
\begin{subfigure}[t]{.46\linewidth}
  \centering
  \includegraphics[height=4.2cm, trim={1.1cm 0 0 0},clip]{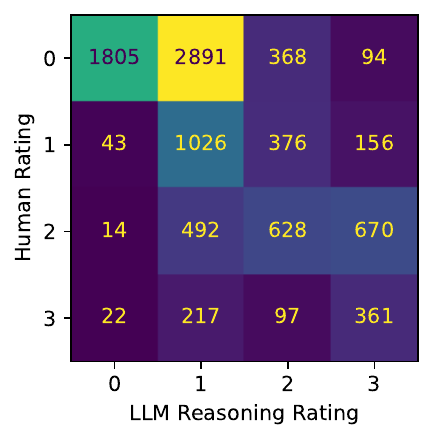}
\end{subfigure}\hfill
\caption{Differences between human and LLM relevance ratings with and without reasoning for GPT 5.1. The LLMs overall are more likely to see relevance in documents, especially with reasoning.}
\label{fig:reasoning_confusion_matrices}
\end{figure}

Since the LLM-RJS can directly estimate prompt relevance under the same definitions as the human annotations, the relevance estimates are compared directly, instead of by the ranking. Figure \ref{fig:reasoning_confusion_matrices} shows the confusion matrices of the non-reasoning and reasoning LLM-RJS relevance rating compared to human relevance ratings. 

We observe reasonable agreement between the LLM-RJS systems and the human annotations, but find that in many cases the LLM-RJS assigns higher relevance ratings than the humans annotations. This is particularly visible in the large number of cases where the human rating has relevance 0 (irrelevant), while the LLM-RJS has relevance 1 (related). This difference increases for the reasoning model, which identifies yet more \mbox{(document, query)} pairs as related. These higher ratings are also reflected in the average relevance ratings, where the average human rating is 0.79, LLM-RJS without reasoning is 1.03, and with reasoning is 1.23. 

Particularly interesting is that the reasoning configuration, which would be expected to perform better, has double the incidences where human ratings identify a pair as irrelevant (0), while the LLM-RJS identifies the pair as perfectly relevant. 

Table \ref{tab:examples_reasoning_nuance} shows examples from these cases. Predominantly, the passages are actually highly relevant, but might not be immediately obvious. The ``difference between a mcdouble and a double cheeseburger'' is indeed one slice of cheese instead of two. This may not be obvious to human annotators or short-sighted NERS models that evaluate similarity, but in these cases the reasoning LLM-RJS outperforms the benchmark. 
Overall the results from Experiment 2 show that the addition of reasoning allows the model to identify relevance that is not obvious, beyond the human annotations.

\section{Conclusion}

We proposed that the short-sightedness problem of NERS may be resolved with relevance reasoning LLM-RJS. Similiarity is not always sufficient for relevance. In Experiment 1 we evaluated whether these LLM-RJS outperform NERS on a ranking task, and do not find improvement. In Experiment 2 we added reasoning, and directly compared the LLM-RJS relevance rankings to the human rankings. This showed the the LLM-RJS was able to identify more documents as relevant than the human annotations. This suggests that the human annotations suffer from similar short-sightedness as the NERS systems, explaining why the LLM-RJS do not reach a noticeable improvement compared to NERS when evaluated on NDCG with short-sighted human annotations as ground truth.

\paragraph{Limitations}

The primary limitation to LLM-RJS systems is the added cost compared to embedding models. The price-per-token of top performing LLMs is roughly 10x that of equivalent embedding models. Additionally, for NERS systems implemented at scale, each document only needs to be embedded once even for multiple queries, and all document embeddings can be stored in a vector database for efficient retrieval. The LLM-RJS system demonstrated here requires LLM inference for each (query, document) pair and therefore does not scale well. A solution may be found in CoT re-ranking \cite{wu2025chain}, where an embedding model is used to make a first selection of documents that are similar, and then the reasoning-LLM selects the relevant documents from the subset of similar documents. However, this only addresses the problem where similar documents are not relevant, but would still fail in cases as the example in Figure \ref{fig:concept_relevance_reasoning}, where the document is relevant but not similar. 

Additionally, we should consider that the short-sightedness in human annotations is not universal for all datasets. We have shown cases of short-sightedness in this dataset, but through expert annotators, instructions that support further research, and more time for annotations it is possible to construct relevance annotations that do not suffer from short-sightedness. Since LLM-RJS models do not need large training datasets, efforts can be focused on only creating test sets.

\paragraph{Future work}
The presented results suggest, but do not definitively show that LLM-RJS models can outperform human annotations. To establish the performance difference between human relevance and LLM-RJS relevance re-annotation may be done where trained annotators compare existing labels and LLM-RJS relevance judgements and establish which is correct. This can be supported by additionally providing the reasoning chain from the LLM-RJS, to provide argumentation for the judgment. 

We have established in this instance of an Information Retrieval a mismatch between similarity and relevance, and show this as a nearsightedness-bottleneck for embeddings. However, for a variety of different Information Retrieval tasks, a variety of retrieval requirements exist beyond `relevance'. In tasks further conditions apply such as preferring high-quality documents, while in other tasks specifically similarity might be preferred over relevance (e.g. for detecting copyright infringement). With advanced Information Retrieval capabilities of LLM-RJS models, soft requirements can be considered, and there are more possibilities to deviate from `similarity' or requirements that can be quantified. Evaluating LLM-RJS against human annotations still appears to be the standard \cite{arabzadeh2025benchmarking}, and efforts are needed to re-specify what the (sub)tasks of Information Retrieval problems are, and how LLM-RJS models can be evaluated. 

In conclusion, we found that the short-sightedness motivation presented in the illustrative Figure \ref{fig:concept_relevance_reasoning} is a valid limitation of NERS models that can be addressed by LLM-RJS. However, as shown with the examples of short-sighted annotations in Table \ref{tab:examples_reasoning_nuance}, human annotations suffer from the same short-sightedness limitation, they present as false-positives for a confusion matrix in Figure \ref{fig:reasoning_confusion_matrices}. Therefore, under a standard annotated benchmark evaluation as shown in Figure \ref{fig:violin_plots}, they do not demonstrably outperform NERS, while they still could be giving better relevance judgments.

\paragraph{Declaration on Generative AI:}  
For this paper, OpenAi's GPT-5, GPT-5.1 and GPT-5.2 models were used for minor reformulations and spelling checking, and LaTeX formatting.
\bibliography{./custom}

\begin{thebibliography}{38}
\providecommand{\natexlab}[1]{#1}

\bibitem[{Abbasiantaeb et~al.(2024)Abbasiantaeb, Meng, Azzopardi, and Aliannejadi}]{abbasiantaeb2024can}
Zahra Abbasiantaeb, Chuan Meng, Leif Azzopardi, and Mohammad Aliannejadi. 2024.
\newblock Can we use large language models to fill relevance judgment holes?
\newblock \emph{arXiv preprint arXiv:2405.05600}.

\bibitem[{Ackley et~al.(1985)Ackley, Hinton, and Sejnowski}]{ackley1985learning}
David~H Ackley, Geoffrey~E Hinton, and Terrence~J Sejnowski. 1985.
\newblock A learning algorithm for boltzmann machines.
\newblock \emph{Cognitive science}, 9(1):147--169.

\bibitem[{Alaofi et~al.(2024)Alaofi, Thomas, Scholer, and Sanderson}]{alaofi2024llms}
Marwah Alaofi, Paul Thomas, Falk Scholer, and Mark Sanderson. 2024.
\newblock Llms can be fooled into labelling a document as relevant: best caf{\'e} near me; this paper is perfectly relevant.
\newblock In \emph{Proceedings of the 2024 Annual International ACM SIGIR Conference on Research and Development in Information Retrieval in the Asia Pacific Region}, pages 32--41.

\bibitem[{{Anthropic}(2025)}]{anthropic_sonnet45}
{Anthropic}. 2025.
\newblock {Claude Sonnet 4.5}.
\newblock \url{https://www.anthropic.com/news/claude-sonnet-4-5}.
\newblock Accessed: 2026-01.

\bibitem[{Arabzadeh and Clarke(2025)}]{arabzadeh2025benchmarking}
Negar Arabzadeh and Charles~LA Clarke. 2025.
\newblock Benchmarking llm-based relevance judgment methods.
\newblock In \emph{Proceedings of the 48th International ACM SIGIR Conference on Research and Development in Information Retrieval}, pages 3194--3204.

\bibitem[{Bailey et~al.(2008)Bailey, Craswell, Soboroff, Thomas, de~Vries, and Yilmaz}]{10.1145/1390334.1390447}
Peter Bailey, Nick Craswell, Ian Soboroff, Paul Thomas, Arjen~P. de~Vries, and Emine Yilmaz. 2008.
\newblock \href {https://doi.org/10.1145/1390334.1390447} {Relevance assessment: are judges exchangeable and does it matter}.
\newblock In \emph{Proceedings of the 31st Annual International ACM SIGIR Conference on Research and Development in Information Retrieval}, SIGIR '08, page 667–674, New York, NY, USA. Association for Computing Machinery.

\bibitem[{Brown et~al.(2020)Brown, Mann, Ryder, Subbiah, Kaplan, Dhariwal, Neelakantan, Shyam, Sastry, Askell et~al.}]{brown2020language}
Tom Brown, Benjamin Mann, Nick Ryder, Melanie Subbiah, Jared~D Kaplan, Prafulla Dhariwal, Arvind Neelakantan, Pranav Shyam, Girish Sastry, Amanda Askell, and 1 others. 2020.
\newblock Language models are few-shot learners.
\newblock \emph{Advances in neural information processing systems}, 33:1877--1901.

\bibitem[{Devlin et~al.(2019)Devlin, Chang, Lee, and Toutanova}]{devlin2019bert}
Jacob Devlin, Ming-Wei Chang, Kenton Lee, and Kristina Toutanova. 2019.
\newblock Bert: Pre-training of deep bidirectional transformers for language understanding.
\newblock In \emph{Proceedings of the 2019 conference of the North American chapter of the association for computational linguistics: human language technologies, volume 1 (long and short papers)}, pages 4171--4186.

\bibitem[{Faggioli et~al.(2023)Faggioli, Dietz, Clarke, Demartini, Hagen, Hauff, Kando, Kanoulas, Potthast, Stein et~al.}]{faggioli2023perspectives}
Guglielmo Faggioli, Laura Dietz, Charles~LA Clarke, Gianluca Demartini, Matthias Hagen, Claudia Hauff, Noriko Kando, Evangelos Kanoulas, Martin Potthast, Benno Stein, and 1 others. 2023.
\newblock Perspectives on large language models for relevance judgment.
\newblock In \emph{Proceedings of the 2023 ACM SIGIR international conference on theory of information retrieval}, pages 39--50.

\bibitem[{Gao et~al.(2021)Gao, Yao, and Chen}]{gao2021simcse}
Tianyu Gao, Xingcheng Yao, and Danqi Chen. 2021.
\newblock Simcse: Simple contrastive learning of sentence embeddings.
\newblock \emph{arXiv preprint arXiv:2104.08821}.

\bibitem[{{Google}(2025)}]{google_gemini25}
{Google}. 2025.
\newblock {Gemini 2.5 Pro}.
\newblock \url{https://docs.cloud.google.com/vertex-ai/generative-ai/docs/models/gemini/2-5-pro}.
\newblock Accessed: 2026-01.

\bibitem[{Guo et~al.(2025)Guo, Yang, Zhang, Song, Zhang, Xu, Zhu, Ma, Wang, Bi et~al.}]{guo2025deepseek}
Daya Guo, Dejian Yang, Haowei Zhang, Junxiao Song, Ruoyu Zhang, Runxin Xu, Qihao Zhu, Shirong Ma, Peiyi Wang, Xiao Bi, and 1 others. 2025.
\newblock Deepseek-r1: Incentivizing reasoning capability in llms via reinforcement learning.
\newblock \emph{arXiv preprint arXiv:2501.12948}.

\bibitem[{Hofst{\"a}tter et~al.(2021)Hofst{\"a}tter, Lin, Yang, Lin, and Hanbury}]{hofstatter2021efficiently}
Sebastian Hofst{\"a}tter, Sheng-Chieh Lin, Jheng-Hong Yang, Jimmy Lin, and Allan Hanbury. 2021.
\newblock Efficiently teaching an effective dense retriever with balanced topic aware sampling.
\newblock In \emph{Proceedings of the 44th international ACM SIGIR conference on research and development in information retrieval}, pages 113--122.

\bibitem[{Huang et~al.(2013)Huang, He, Gao, Deng, Acero, and Heck}]{huang2013learning}
Po-Sen Huang, Xiaodong He, Jianfeng Gao, Li~Deng, Alex Acero, and Larry Heck. 2013.
\newblock Learning deep structured semantic models for web search using clickthrough data.
\newblock In \emph{Proceedings of the 22nd ACM international conference on Information \& Knowledge Management}, pages 2333--2338.

\bibitem[{Jaech et~al.(2024)Jaech, Kalai, Lerer, Richardson, El-Kishky, Low, Helyar, Madry, Beutel, Carney et~al.}]{jaech2024openai}
Aaron Jaech, Adam Kalai, Adam Lerer, Adam Richardson, Ahmed El-Kishky, Aiden Low, Alec Helyar, Aleksander Madry, Alex Beutel, Alex Carney, and 1 others. 2024.
\newblock Openai o1 system card.
\newblock \emph{arXiv preprint arXiv:2412.16720}.

\bibitem[{Karpukhin et~al.(2020)Karpukhin, Oguz, Min, Lewis, Wu, Edunov, Chen, and Yih}]{karpukhin2020dense}
Vladimir Karpukhin, Barlas Oguz, Sewon Min, Patrick~SH Lewis, Ledell Wu, Sergey Edunov, Danqi Chen, and Wen-tau Yih. 2020.
\newblock Dense passage retrieval for open-domain question answering.
\newblock In \emph{EMNLP (1)}, pages 6769--6781.

\bibitem[{Kazi et~al.(2025)Kazi, Khoja, and Daud}]{kazi2025bridging}
Samreen Kazi, Shakeel Khoja, and Ali Daud. 2025.
\newblock Bridging the gap: A survey of document retrieval techniques for high-resource and low-resource languages.
\newblock \emph{Computer Science Review}, 57:100756.

\bibitem[{Kusupati et~al.(2022)Kusupati, Bhatt, Rege, Wallingford, Sinha, Ramanujan, Howard-Snyder, Chen, Kakade, Jain, and Farhadi}]{NEURIPS2022_c32319f4}
Aditya Kusupati, Gantavya Bhatt, Aniket Rege, Matthew Wallingford, Aditya Sinha, Vivek Ramanujan, William Howard-Snyder, Kaifeng Chen, Sham Kakade, Prateek Jain, and Ali Farhadi. 2022.
\newblock \href {https://proceedings.neurips.cc/paper_files/paper/2022/file/c32319f4868da7613d78af9993100e42-Paper-Conference.pdf} {Matryoshka representation learning}.
\newblock In \emph{Advances in Neural Information Processing Systems}, volume~35, pages 30233--30249. Curran Associates, Inc.

\bibitem[{Lee et~al.(2025)Lee, Chen, Dua, Cer, Shanbhogue, Naim, {\'A}brego, Li, Chen, Vera et~al.}]{lee2025gemini}
Jinhyuk Lee, Feiyang Chen, Sahil Dua, Daniel Cer, Madhuri Shanbhogue, Iftekhar Naim, Gustavo~Hern{\'a}ndez {\'A}brego, Zhe Li, Kaifeng Chen, Henrique~Schechter Vera, and 1 others. 2025.
\newblock Gemini embedding: Generalizable embeddings from gemini.
\newblock \emph{arXiv preprint arXiv:2503.07891}.

\bibitem[{MacAvaney and Soldaini(2023)}]{macavaney2023one}
Sean MacAvaney and Luca Soldaini. 2023.
\newblock One-shot labeling for automatic relevance estimation.
\newblock In \emph{Proceedings of the 46th International ACM SIGIR Conference on Research and Development in Information Retrieval}, pages 2230--2235.

\bibitem[{{Meta}(2024{\natexlab{a}})}]{meta_llama31}
{Meta}. 2024{\natexlab{a}}.
\newblock {LLaMA 3.1 Models}.
\newblock \url{https://ollama.com/library/llama3.1}.
\newblock Accessed: 2026-01.

\bibitem[{{Meta}(2024{\natexlab{b}})}]{meta_llama32}
{Meta}. 2024{\natexlab{b}}.
\newblock {LLaMA 3.2 Models}.
\newblock \url{https://ollama.com/library/llama3.2}.
\newblock Accessed: 2026-01.

\bibitem[{Mikolov et~al.(2013)Mikolov, Chen, Corrado, and Dean}]{mikolov2013efficient}
Tomas Mikolov, Kai Chen, Greg Corrado, and Jeffrey Dean. 2013.
\newblock Efficient estimation of word representations in vector space.
\newblock \emph{arXiv preprint arXiv:1301.3781}.

\bibitem[{Nussbaum et~al.(2024)Nussbaum, Morris, Duderstadt, and Mulyar}]{nussbaum2024nomic}
Zach Nussbaum, John~X Morris, Brandon Duderstadt, and Andriy Mulyar. 2024.
\newblock Nomic embed: Training a reproducible long context text embedder.
\newblock \emph{arXiv preprint arXiv:2402.01613}.

\bibitem[{{Ollama}(2024)}]{ollama_gptoss}
{Ollama}. 2024.
\newblock {GPT-OSS Models}.
\newblock \url{https://ollama.com/library/gpt-oss}.
\newblock Accessed: 2026-01.

\bibitem[{{OpenAI}(2025)}]{openai_gpt51}
{OpenAI}. 2025.
\newblock {GPT-5.1 Model}.
\newblock \url{https://openai.com/index/gpt-5-1/}.
\newblock Accessed: 2026-01.

\bibitem[{Peeperkorn et~al.(2024)Peeperkorn, Kouwenhoven, Brown, and Jordanous}]{peeperkorn2024temperature}
Max Peeperkorn, Tom Kouwenhoven, Dan Brown, and Anna Jordanous. 2024.
\newblock Is temperature the creativity parameter of large language models?
\newblock \emph{arXiv preprint arXiv:2405.00492}.

\bibitem[{Pennington et~al.(2014)Pennington, Socher, and Manning}]{pennington2014glove}
Jeffrey Pennington, Richard Socher, and Christopher~D Manning. 2014.
\newblock Glove: Global vectors for word representation.
\newblock In \emph{Proceedings of the 2014 conference on empirical methods in natural language processing (EMNLP)}, pages 1532--1543.

\bibitem[{Robertson et~al.(2009)Robertson, Zaragoza et~al.}]{robertson2009probabilistic}
Stephen Robertson, Hugo Zaragoza, and 1 others. 2009.
\newblock The probabilistic relevance framework: Bm25 and beyond.
\newblock \emph{Foundations and Trends{\textregistered} in Information Retrieval}, 3(4):333--389.

\bibitem[{Salton(1965)}]{salton1965evaluation}
Gerard Salton. 1965.
\newblock The evaluation of automatic retrieval procedures—selected test results using the smart system.
\newblock \emph{American Documentation}, 16(3):209--222.

\bibitem[{Salton et~al.(1975)Salton, Wong, and Yang}]{salton1975vector}
Gerard Salton, Anita Wong, and Chung-Shu Yang. 1975.
\newblock A vector space model for automatic indexing.
\newblock \emph{Communications of the ACM}, 18(11):613--620.

\bibitem[{Trabelsi et~al.(2021)Trabelsi, Chen, Davison, and Heflin}]{trabelsi2021neural}
Mohamed Trabelsi, Zhiyu Chen, Brian~D Davison, and Jeff Heflin. 2021.
\newblock Neural ranking models for document retrieval.
\newblock \emph{Information Retrieval Journal}, 24(6):400--444.

\bibitem[{Voorhees et~al.(2020)Voorhees, Craswell, Mitra, Campos, and Yilmaz}]{trec2019}
Ellen Voorhees, Nick Craswell, Bhaskar Mitra, Daniel Campos, and Emine Yilmaz. 2020.
\newblock \href {https://tsapps.nist.gov/publication/get_pdf.cfm?pub_id=930061} {Overview of the trec 2019 deep learning track}.

\bibitem[{Wang and Dong(2020)}]{wang2020measurement}
Jiapeng Wang and Yihong Dong. 2020.
\newblock Measurement of text similarity: a survey.
\newblock \emph{Information}, 11(9):421.

\bibitem[{Wang et~al.(2013)Wang, Wang, Li, He, and Liu}]{wang2013theoretical}
Yining Wang, Liwei Wang, Yuanzhi Li, Di~He, and Tie-Yan Liu. 2013.
\newblock A theoretical analysis of ndcg type ranking measures.
\newblock In \emph{Conference on learning theory}, pages 25--54. PMLR.

\bibitem[{Wei et~al.(2022)Wei, Wang, Schuurmans, Bosma, Xia, Chi, Le, Zhou et~al.}]{wei2022chain}
Jason Wei, Xuezhi Wang, Dale Schuurmans, Maarten Bosma, Fei Xia, Ed~Chi, Quoc~V Le, Denny Zhou, and 1 others. 2022.
\newblock Chain-of-thought prompting elicits reasoning in large language models.
\newblock \emph{Advances in neural information processing systems}, 35:24824--24837.

\bibitem[{Wu et~al.(2025)Wu, Zhou, Chen, Zhang, and Mok}]{wu2025chain}
Shangrong Wu, Yanghong Zhou, Yang Chen, Feng Zhang, and PY~Mok. 2025.
\newblock Chain-of-thought re-ranking for image retrieval tasks.
\newblock \emph{arXiv preprint arXiv:2509.14746}.

\bibitem[{Xing et~al.(2023)Xing, Ma, Qian, Lin, and Hu}]{xing2023query}
W~Xing, Guangyuan Ma, Wanhui Qian, Zijia Lin, and Songlin Hu. 2023.
\newblock Query-as-context pre-training for dense passage retrieval.
\newblock In \emph{Proceedings of the 2023 Conference on Empirical Methods in Natural Language Processing}, pages 1906--1916.

\end{thebibliography}

\newpage

\onecolumn

\appendix
\section{Prompt}\label{ap:appendixPrompt}
\label{Prompt}
The prompt that was used for all LLMs was:
\begin{small}
\begin{Verbatim}[breaklines]

You are an automatic passage relevance judge for information retrieval.
Your task is to rate how relevant a single passage is to a given query.

Follow these instructions:

- Carefully read the query and the passage.
- Decide how relevant the passage is to the query only, ignoring any external knowledge you might have.
- Assign exactly one integer relevance score using this 4-point scale:

3 - Perfectly relevant:
The passage is clearly dedicated to the query and contains the exact answer or all key information needed to answer it. No important part of the answer is missing.

Example:
Query: "What is chocolate?"
Passage: "Chocolate is a food made by roasting and grinding cacao seeds."
Score: 3

2 - Highly relevant:
The passage contains some answer to the query, but the answer may be partially unclear, incomplete, indirect, or mixed with a lot of extraneous information. Still, it provides clear help in answering the query.

Example:
Query: "Labrador breed"
Passage: "Labradors are friendly dogs often used as guide animals."
Score: 2
(Passage is about Labradors but doesn't define the breed fully.)

1 - Related:
The passage is about a topic clearly related to the query, but it does not actually answer the query. It may provide background, context, or tangential information only.

Example:
Query: "What is chocolate?"
Passage: "Cacao trees grow in tropical 
climates."
Score: 1 (Related ingredient, but no definition of chocolate.)

0 - Irrelevant:
The passage has nothing to do with the query or is only extremely weakly related, and it does not help answer the query at all.

Example:
Query: "Labrador breed"
Passage: "The Eiffel Tower is in Paris."
Score: 0



Output format:

You must always return a single integer score between 0 and 3 and NOTHING else.
\end{Verbatim}
\end{small}
\newpage

\section{Genuine False Positives in GPT 5.1 with reasoning}\label{ap:false_positives}

\begin{table*}[h]
\centering
\footnotesize
\setlength{\tabcolsep}{3pt}
\renewcommand{\arraystretch}{1.1}
\begin{tabularx}{\textwidth}{@{}
>{\RaggedRight\arraybackslash}p{0.18\textwidth}
>{\RaggedRight\arraybackslash}X
c c c c
@{}}
\toprule
\textbf{Query} & \textbf{Passage} & \textbf{Human} & \textbf{Reasoning} & \textbf{No Reasoning} & \textbf{NERS} \\
\midrule

exons definition biology & [...]recursor mRNA (pre-mRNA) is transcribed by RNA polymerase II in the nucleus. pre-mRNA is then processed by splicing to remove introns, leaving the exons in the mature messenger RNA (mRNA). [...] & 0 & 3 & 2& 0.67\\
is cdg airport in main paris & CDG may refer to: [...] 2  Charles de Gaulle, French general and politician.  IATA code for Charles de Gaulle Airport, in Paris, France. & 0  & 3 & 2 & 0.52\\

lps laws definition &	Lipopolysaccharides (LPS), also known as lipoglycans and endotoxins, are large molecules consisting of a lipid [...] & 0 & 3 & 1 & 0.66\\

lps laws definition	& Bacterial lipopolysaccharides (LPS) are the major outer surface membrane components present in almost all Gram-negative bacteria [...] & 0 & 3 & 1 & 0.69\\

why did the us volunterilay enter ww1 &	The United States in World War I declared war on the German Empire on April 6, 1917. The U.S. was an independent power and did not officially join the Allies. It closely cooperated with them militarily but acted alone in diplomacy.n January 1917 Germany resumed unrestricted submarine warfare. The German Foreign minister, Arthur Zimmermann invited revolution-torn Mexico to join the war as Germany's ally against the United States in the Zimmermann Telegram. & 0 & 3 & 2 & 0.56\\

\bottomrule
\end{tabularx}
\caption{%
The 5 false positives where GPT 5.1 with reasoning gave a relevance rating of 3, while the human annotations gave a relevance of 0. While many passages are partially relevant, we believe they do not fully answer the query and should receive a lower rating. All other 89 `false positives' are due to mistakes in the annotations, not mistakes of the LLM-RJS. %
}
\label{tab:examples_genuine_false_positives}
\end{table*}

\section{Reasoning experiment with GPT-OSS}\label{ap:oss-reasoning}

\begin{figure}[ht]
\centering
\begin{subfigure}[t]{.54\linewidth}
  \centering
  \includegraphics[height=6cm]{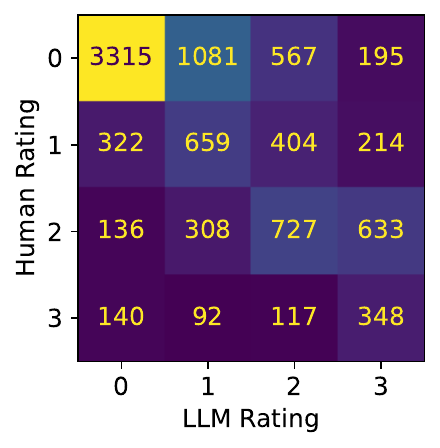}
\end{subfigure}\hfill
\begin{subfigure}[t]{.46\linewidth}
  \centering
  \includegraphics[height=6cm, trim={1.1cm 0 0 0},clip]{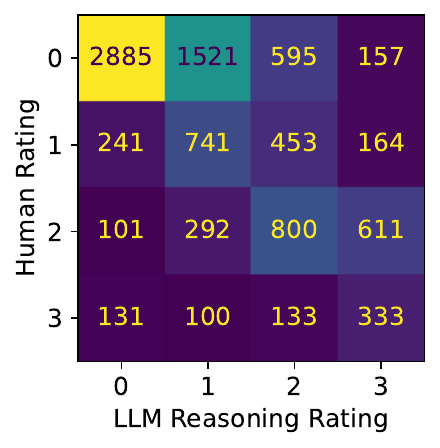}
\end{subfigure}\hfill
\caption{Differences between human and LLM relevance ratings with and without reasoning for GPT 5.1. The LLMs overall are more likely to see relevance in documents, especially with reasoning.}
\label{fig:reasoning_confusion_matrices_oss}
\end{figure}

In Figure \ref{fig:reasoning_confusion_matrices_oss} we reproduce Experiment 2 with an open-source reasoning model. Again we see that the model is more likely to give higher relevance ratings than the humans, and that this effect is larger for the reasoning model. Compared to GPT 5.1, we do see an increase in false-negatives (lower triangular) suggesting that this model might also have generally increased errors. 

\FloatBarrier

\section{Examples of LLM-RJS failure}
\begin{table}[H]
\centering
\footnotesize
\setlength{\tabcolsep}{4pt}
\renewcommand{\arraystretch}{1.15}

\begin{tabularx}{\textwidth}{@{}
  >{\RaggedRight\arraybackslash}p{0.12\textwidth}
  >{\RaggedRight\arraybackslash}p{0.20\textwidth}
  Y
  Y
@{}}
\toprule
\textbf{LLM} & \textbf{Query} & \textbf{Passage} & \textbf{LLM output} \\
\midrule
Sonnet 4.5 &
anthropological definition of environment &
Man and environment reciprocal relationship. The relationship between humans and environment has varied from the early periods of human settlement on the earth to the present day. The relationship between environment and human beings has also being varying from place to place at any given period of time. &
1

The passage discusses the relationship between humans and environment from an anthropological perspective, which is related to the query topic. However, it does not actually provide a definition of ``environment'' in anthropological terms. It only describes how the human--environment relationship varies across time and space, which is contextual information rather than a definition.
\\

\addlinespace[0.6em]

llama3.2:1b &
cost of interior concrete flooring &
1 To have a professional complete the epoxy flooring, the price can vary anywhere from \$5 to \$13 per square foot. 2  An average two car garage can cost anywhere from \$2,000 to \$5,000 to complete.3  To get an exact price for an installation job, consider using free services such as HomeAdvisor.com. For example, a water-based epoxy paint can retail for \$20 to \$40. 2  To have a professional complete the epoxy flooring, the price can vary anywhere from \$5 to \$13 per square foot. 3  An average two car garage can cost anywhere from \$2,000 to \$5,000 to complete. &
You must assign exactly one integer relevance score.

Score: 1
\\

\addlinespace[0.6em]

llama3.1:8b &
do goldfish grow &
The dolphin fish is the fastest growing wild fish species known to man. When the conditions are right, the dolphin fish can grow as fast as 1.3 to 2.7 inches in a week!That means that in about 1 year, the fish can grow to four feet long and could be as heavy as 40 pounds. Studies have shown that females rarely go over 40 pounds while males can grow to as much as 60 pounds. As a fast-growing fish species, it is also among those which mature early. In fact, dolphin fish are sexually mature in as early as 3 months, some even maturing at just 14 inches long.&
You want me to rate the relevance of the passage to the query ``do goldfish grow".

After carefully reading both the query and the passage, I assign a score of:

Score: 2

The passage provides clear information about the growth rate of dolphin fish, including their potential size and weight. The passage also mentions that studies have shown females rarely exceed 40 pounds while males can reach up to 60 pounds, which is relevant to the query.

However, the passage does not directly answer whether goldfish grow or not. It provides general information about dolphin fish's growth rate but does not provide any direct evidence or examples of goldfish growing. Therefore, I assign a score of 2 instead of 3, indicating that the passage is related to the query but does not provide a definitive answer.

\\

\addlinespace[0.6em]

\bottomrule
\end{tabularx}

\caption{Examples of outputs of models that failed to comply to the prompt. The models fail to only give the relevance score and provide additional justification.}
\label{tab:failed_examples}
\end{table}

\end{document}